\documentclass[pre,twocolumn,a4paper,showpacs]{revtex4}
\usepackage{amssymb}
\usepackage{graphicx}
\usepackage{epsfig}
\begin{document}

\title{The media effect in Axelrod's model explained}

\author{Lucas R. Peres and Jos\'e F. Fontanari}
\affiliation{Instituto de F\'{\i}sica de S\~ao Carlos,
  Universidade de S\~ao Paulo,
  Caixa Postal 369, 13560-970 S\~ao Carlos, S\~ao Paulo, Brazil}


\begin{abstract}
 We revisit the problem of introducing an external  global field -- the mass
media -- in  Axelrod's model of social dynamics, where in addition to their nearest neighbors,
the agents can interact with a virtual neighbor whose
cultural features are fixed from the outset. The  finding that this apparently
homogenizing field actually increases the cultural diversity has been considered a 
puzzle since the phenomenon was first reported more than a decade ago.
Here  we offer  a simple explanation for it,  which  is based on the pedestrian 
observation that Axelrod's model exhibits more cultural diversity, i.e., more distinct 
cultural domains,  when the agents are allowed to interact solely with the media field than
when they can interact with their neighbors as well. In this perspective,  it is the local 
homogenizing interactions that work towards  making the
absorbing configurations less fragmented as compared with the extreme  situation in which the agents
interact with the media only.
\end{abstract}

\maketitle

\section{Introduction}

Although computational social science -- a branch of computer science that aims at predicting the behavior of groups as well as 
exploring ways of influencing that behavior -- is currently in vogue  in  the scientific media (see, e.g., \cite{Nature_11}), it has a
long record in the computational physics community \cite{Toral_07, Galam_08,Castellano_09}. In fact, physicists have offered a variety of
models of social influence and opinion formation  (see, e.g.,  \cite{Lewenstein_92,Sznajd_00,Galam_05,Carletti_06}), but they seem to have
adopted as a paradigm a model proposed by 
a political  scientist --  Axelrod's model  \cite{Axelrod_97} -- which exhibits the right  balance between simplicity and realism, and so succeeds 
in boiling down a collective social phenomenon  to its functional essence \cite{Goldstone_05}. 
In addition, Axelrod's model has served as  inspiration to the proposal of   heuristics that use social interactions to solve combinatorial 
optimization problems \cite{Kennedy_98,Swarm,Fontanari_10}.

Axelrod's   model  was introduced to explore the mechanisms behind the persistence of
cultural and opinion differences in a society, given  that the interactions between people tend, on the average,  to make
them more alike in their beliefs and attitudes. Hinging on  a few reasonable assumptions, Axelrod's model  offers  a
nontrivial answer to that issue. In that model, an agent  is represented by a string  of
cultural features, where each feature can adopt a certain number of distinct traits. The interaction between any two agents takes place
with probability proportional to their cultural similarity, i.e., proportional to the number of traits they have in common.
The result of such interaction is the increase of the similarity between the two agents, as one of them modifies 
a previously  distinct trait to match that of its partner. In spite of this homogenizing 
assumption, the model exhibits global polarization, i.e., a stable multicultural regime
\cite{Axelrod_97}. Subsequent analysis of Axelrod's model by the statistical physics community has revealed a rich 
dynamic behavior with a nonequilibrium  phase transition separating the global polarization regime
from the homogeneous regime, where a single culture dominates the entire population 
\cite{Castellano_00, Klemm_03a,Barbosa_09}. 

Another counterintuitive result derived from the analysis of Axelrod's model is that the introduction
of an external  global  media (i.e., it is the same for all agents) aiming at influencing the agents’ opinions actually
favors polarization \cite{Shibanai_01}. This result  conflicts with the view that mass media,
such as newspapers and television, are  instruments  to control people's
opinions and so homogenize society \cite{Lazarsfeld_48}. The effect of the mass media in Axelrod's model has been extensively investigated
(see, e.g., \cite{Avella_05,Avella_06,Mazzitello_07,Candia_08}), but no first-principles explanation for it was offered. 
In fact, much of that research has focused on the search for a threshold on the intensity of the media influence such that above
that threshold, the population is culturally polarized and below it, the population is  homogeneous. More recently, it was argued that this threshold is 
an artifact of finite lattices, and that even a vanishingly small media influence is sufficient to produce   cultural
polarization   in a region of the parameter space  where  the homogeneous regime is dominant  in the absence of the media \cite{Peres_10a}.

Here we offer a  simple explanation for the media effect in Axelrod's model,
 which is based on  viewing the problem 
from a new perspective: rather than focusing on the counterintuitive domain-fragmenting effect of increasing the media strength,
we centre at the trite homogenizing effect of the interactions between agents, which become more important as the media strength 
is weakened. We conclude then that   it is the local  homogenizing interactions that work towards  making the
absorbing configurations less heterogeneous as compared with the extreme  situation in which the agents
can interact with the media only.

\section{Model}
In the original Axelrod's model \cite{Axelrod_97}, each agent is characterized by a set of $F$ cultural features which can take on
$q$ distinct values. The agents are fixed in the sites of a square lattice of linear size $L$ with free boundary conditions
(i.e., agents in the corners of the lattice interact with two neighbors, agents in the sides  with three, 
and agents in the bulk with four nearest neighbors).  These were the boundary conditions used in the seminal paper by
Axelrod \cite{Axelrod_97}, which we adopt here because they greatly facilitate the implementation of Hoshen and Kopelman algorithm for
counting the number of clusters in a lattice \cite{Stauffer_92}. 
The initial configuration is completely random with the features of each agent given by  
random integers drawn uniformly between $1$ and $q$. At each time we pick an agent at random
(this is the target agent) as well as one of its neighbors. These two 
agents interact with probability equal to  their cultural similarity, defined as the fraction of 
common cultural features. An interaction consists of selecting at random one of the distinct features, and making the
selected feature of the target agent equal  to its neighbor's corresponding trait. This procedure is repeated until 
the system is frozen into an absorbing configuration. 

The introduction of a global media in the standard model follows the ingenious suggestion 
 of adding   a virtual  agent which interacts with all agents in the lattice and whose cultural traits reflect the media 
message \cite{Shibanai_01}. In the original version,  each 
cultural feature of the virtual agent has the trait which is the most common in the population --
the consensus opinion. To speed up the simulations,
here we choose to keep the media message
fixed from the outset, so it really models some alien influence impinging on the population.
Explicitly, we generate the culture vector of the virtual agent at random and keep it fixed
during the dynamics \cite{Avella_05,Avella_06}. The interaction of the media (virtual agent) with the real agents is governed by
the control parameter
$p \in \left [ 0,1 \right ] $, which may be interpreted as a measure of the strength of the media influence. 
As in the original Axelrod's
model, we begin by choosing  a target agent at random, but now it  can interact with the media with probability $p$ 
or with its neighbors with probability $1-p$.  Since we have defined the media as a virtual  agent, 
the interaction follows exactly the same rules as before.  The original model is recovered by setting $p=0$. 

The simulation of the dynamics of  Axelrod's model can be greatly accelerated by introducing a list of  active agents \cite{Barbosa_09}. 
An active agent is an agent 
that has at least one feature in common  and at least one feature distinct with at least one of its neighbors, included the virtual
agent (media).
Clearly, since only active agents can change their cultural features, it is more efficient  to select the
target agent randomly from the list of active agents rather than from the entire lattice. Note that the
randomly selected neighbor of the target agent may not necessarily be an active agent itself. In the case that
the cultural features of the target agent are modified by the interaction with its neighbor, we 
need to re-examine the active/inactive status of the target agent as well as of all its neighbors so as
to update the list of active agents.  The dynamics is frozen when the list of active agents is empty.

Because the cultural features of the virtual agent are fixed a priori, there are three distinct types of absorbing states  of 
the dynamics.  These absorbing states determine the stationary (or frozen) regimes of the model.
First, there is the homogeneous regime where all agents are identical to the media.  Second, there are $\left ( q-1 \right )^F$  distinct 
homogeneous regimes in which all $F$ cultural traits of the agents are distinct from those of the media. However, we will show below that this
type of absorbing configuration is absent for large lattices.  Third, there is a
heterogeneous regime characterized by the co-existence of domains of  agents whose cultural traits are  
identical to the media with domains of agents  whose traits are all distinct from the media.

As pointed out before, the puzzling phenomenon reported first  by Shibanai et al. \cite{Shibanai_01} and then confirmed by many
other authors \cite{Avella_05,Avella_06,Mazzitello_07,Candia_08,Peres_10a} is that for a set of control parameters, say $F=5$ and $q=5$, 
the homogeneous regime in the absence of media is replaced by a heterogeneous regime whenever the media is turned on (i.e., for $p >0$). 
This goes against the intuitive idea that the effect of the media is to make the culture more homogeneous. 
Studies based on  lattices of small sizes (typically $L =40$) have suggested the existence of
 a threshold value $p_c$  on the intensity of the media influence such that for $p > p_c$ the stationary regime is heterogeneous, whereas 
it is homogeneous for $p \leq p_c$ \cite{Avella_05,Avella_06,Mazzitello_07,Candia_08}. (We recall that the parameters $q$ and $F$ are chosen such that the population is
culturally homogeneous for $p=0$.) This  finding was questioned in a recent study which suggested  that $p_c$  is 
an artifact of finite lattices and hence that even a vanishingly small media influence is sufficient to destabilize the culturally
homogeneous regime \cite{Peres_10a}. However, the problem with that study was that the virtual agent (media) did not take part in
the determination  of the active status of the agents. As we will argue below, correction of this point does not change the claim that
$p_c \to 0$ when $L \to \infty$.

\section{Simulation Results}

\begin{figure}
\centerline{\epsfig{width=0.52\textwidth,file=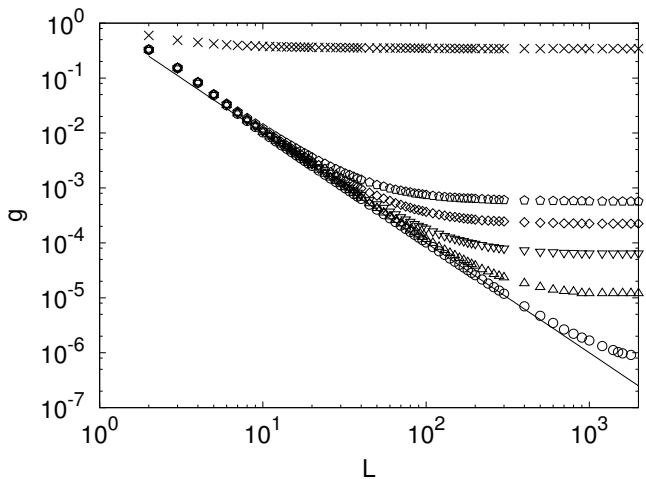}}
\par
\caption{Logarithmic plot of the ratio $g$ between the number of domains and the lattice area  as function of the linear size $L$ of the lattice for (top to
bottom) $p = 1, 0.05, 0.04, 0.03, 0.02$ and $0.01$. The solid straight line is $1/L^2$, which corresponds
to the value of $g$ in the uniform regime. The parameters are $F = 5$ and $q = 5$.}
\label{fig:1}
\end{figure}

We consider first  the   ratio $g \in \left [ 0, 1 \right ] $ between the
number of clusters or cultural domains $S$ and the lattice area $L^2$ (see fig.\ \ref{fig:1}).  
A cluster or domain is simply a
bounded region of uniform culture. Two or more cultural domains characterized by the same culture are counted separately in
the calculation of $S$. For the parameter set $F=q=5$, analysis of  small lattices indicates a threshold value $p_c \approx 0.03$
\cite{Avella_05,Avella_06}, but the results exhibited in fig.\  \ref{fig:1} indicate unequivocally that the heterogeneous regime
displaces the homogeneous one  even for very small values of the media strength as the lattice size increases. This figure is
virtually indistinguishable from that obtained in the case the  media  does not take part in the determination of the
 active status of the agents \cite{Peres_10a}.  In this,  as well as  in the next figures of this paper,
the error bars are smaller or at most equal to the symbol sizes.
We note that there are many more cultural domains in the case
 the local interactions are turned off (i.e., $p=1$) than in any other situation.

\begin{figure}
\centerline{\epsfig{width=0.52\textwidth,file=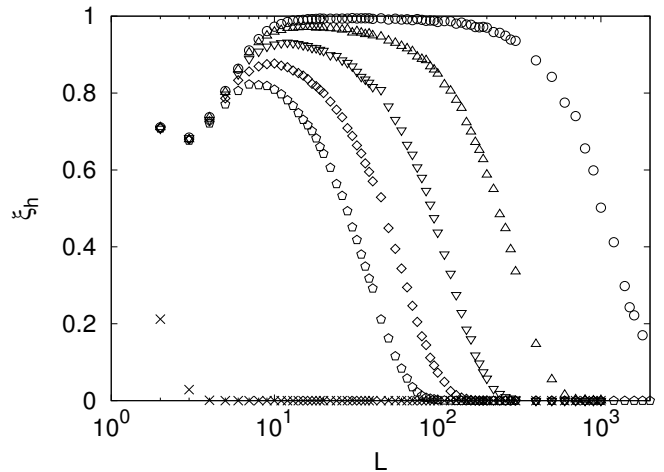}}
\caption{Fraction of runs  trapped into homogeneous absorbing configuration $\xi_h$  as function of the linear size $L$ of the lattice for (top to
bottom) $p = 0.01, 0.02, 0.03, 0.04, 0.05$ and $1$. The parameters are $F = 5$ and $q = 5$.}
\label{fig:2}
\end{figure}

\begin{figure}
\centerline{\epsfig{width=0.52\textwidth,file=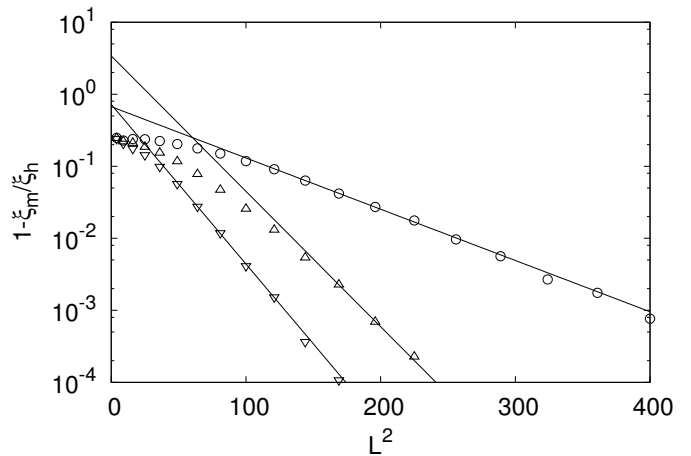}}
\caption{Fraction of the homogeneous absorbing configurations which are distinct from the media  as function of the lattice size $L^2$  for (top to
bottom) $p = 0.01, 0.02 $ and $0.03$. The solid lines are the fittings $a_p \exp \left ( - b_p L^2 \right ) $.
The parameters are $F = 5$ and $q = 5$.}
\label{fig:3}
\end{figure}

A more lively illustration of the effect of increasing the lattice size is offered in fig.\ \ref{fig:2}, which shows the fraction of runs $\xi_h$  which 
 ended up in a homogeneous absorbing configuration. Typically we carried out from $10^3$ to $10^4$ independent runs of the stochastic
 dynamics for a given value of the media strength $p$. This figure highlights   the somewhat odd predominance of the homogeneous 
absorbing configurations for lattices of intermediate size, which has led to the suggestion of a nonzero threshold value $p_c$. 
For $p=1$ and $L > 5$ we find that  the absorbing configurations are always heterogeneous.
In addition, we note that, except for small lattice sizes, the homogeneous absorbing configurations are identical to the media. In fact, we have estimated the fraction of runs
$\xi_m$ which ended up in the media configuration and plotted in  fig.\ \ref{fig:3} 
the fraction of homogeneous absorbing configurations distinct from the media, i.e., the quantity $1 - \xi_m/\xi_h$. 
As shown in the figure, this fraction   vanishes exponentially fast with the increase of the lattice size $L^2$ and so, for large lattices, there is essentially a single    homogeneous 
absorbing configuration -- the one where all agents assume the media cultural traits. 
 
\begin{figure}
\centerline{\epsfig{width=0.52\textwidth,file=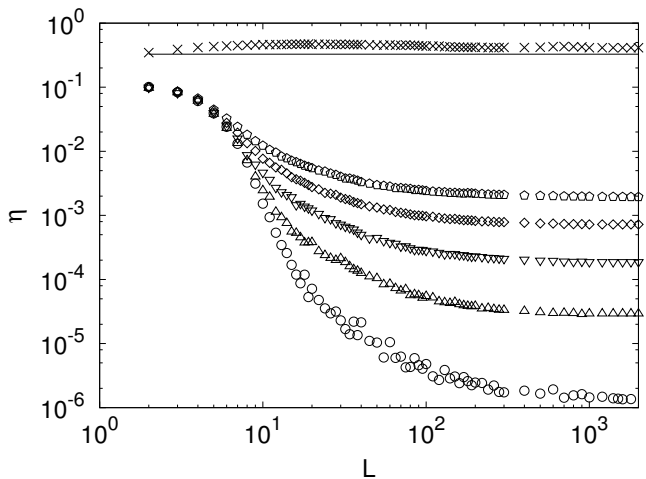}}
\caption{Fraction of the lattice sites that are not part of the largest domain (media)   as function of the linear size $L$ of the lattice for  (top to
bottom) $p = 1, 0.05, 0.04, 0.03, 0.02$ and $0.01$. The horizontal line is $\left ( 1-1/q \right )^F$.The parameters are $F = 5$ and $q = 5$.}
\label{fig:4}
\end{figure}

\begin{figure}
\centerline{\epsfig{width=0.52\textwidth,file=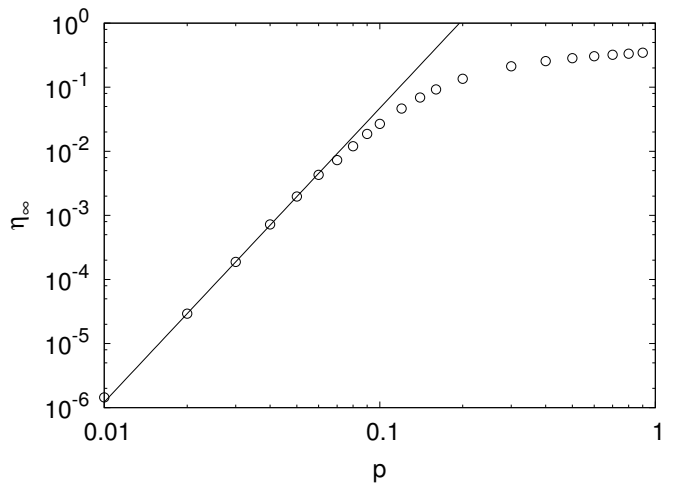}}
\caption{Fraction of the lattice sites that are not part the largest domain (media) in the thermodynamic limit,   $\eta_\infty$,  as function of the media strength $p$.
The solid line is the fitting $\eta_{\infty} = 1820 p^{4.6}$. The parameters are $F = 5$ and $q = 5$.}
\label{fig:5}
\end{figure}

The results exhibited in figs. \ref{fig:1} and \ref{fig:2} merely  reassert the claim that the media promotes cultural diversity, since
the number of cultural domains increases with the strength of the media, measured by the parameter $p$. However,  
this is a rather odd view of the situation. In fact, in all runs and for all  parameter settings considered, we found that the largest domain is  composed of
agents identical to the media. 
In addition, the size of this domain is larger than the sum of the sizes of the other  domains, as shown in
fig.\ \ref{fig:4}, which exhibits the average fraction $\eta$ of the lattice sites (agents) that do not belong to the largest domain. 
Clearly, $\eta = 1 - \left \langle S_{max} \right \rangle/L^2$ where $\left \langle S_{max} \right \rangle$  is the average size of the largest  domain (media) which, together with
$g$, has often been used as an order parameter for Axelrod's model \cite{Klemm_03a,Avella_06}.
We note that although some of the
small domains are also  characterized by the media traits, they count as distinct domains 
as they are disconnected from the largest cluster --  a situation known as cultural diaspora \cite{Greig_02}.  
In agreement with the conclusion drawn from the previous figures, 
the sum of the  sizes of the domains different from the media occupy a finite (i.e., macroscopic)  portion of the lattice in the thermodynamic limit.  
In  this limit, the average size of the domains, 
excluded the largest one,  can be obtained by calculating the ratio $\eta/g$ and yields  about 3 sites, 
regardless of the value of $p$, so they are microscopic domains. This implies that there is only one macroscopic domain in  the system -- the media --
which in the limit $p \to 0$ tends continuously to the size of the entire system. This point  is shown explicitly in fig.\ \ref{fig:5}  where we plot the asymptotic
values of the fraction of lattice sites that are not part of  the largest domain $\eta_\infty$ against the  media strength $p$. 

Figure \ref{fig:4} hints that there is something
trivial about the nature of the cultural domains distinct from the media. 
The are two points which are worth emphasizing here.  First, when the agents are allowed to interact solely with  the fixed external media, the absorbing
configurations are heterogeneous (see fig.\ \ref{fig:2}), since agents whose  $F$ cultural traits   are all distinct from the media  at the onset will be
unaffected by the dynamic rules. For instance,  the horizontal line in  fig.\ \ref{fig:4} gives the
fraction of sites distinct from the media  regardless of the domain organization  in the case the local interactions are turned off ($p=1$),  i.e., $\left ( 1 - 1/q \right ) ^F $. 
Those sites are frozen from the very beginning of the simulation.
Second,  there are many more agents different from the media  for $p=1$ than for $p < 1$, i.e., the local
homogenizing interactions contribute to increase the  number of agents with the media traits. Summing up, due to the non-zero similarity restriction for the occurrence of interactions,
the  media in Axelrod's model
does not  promote uniformity; on the contrary, there are more cultural diversity when the media is the only partner  the agents can interact with than
when local interactions are allowed.

\section{Conclusion}

There is nothing extraordinary about the effect of the media in Axelrod's model, provided we interpret the results summarized in fig. \ref{fig:4}
from top to bottom: when assembling the random initial   configuration a fraction $\left ( 1 - 1/q \right ) ^F$ of the agents have their $F$ cultural
traits distinct from the media and so they form the cultural domains in the case the local interactions are neglected ($p \to 1$). As the media strength $p$
decreases, the local interactions have the chance to flip some of those traits, resulting in the increase of the size of the media cluster. Finally, when
$p \to 0$ the media domain takes over the entire lattice, leading to a unique homogeneous absorbing configuration. For $p=0$ the homogeneous absorbing
configuration is not unique -- there are $q^F$ different possibilities -- but, as in the ferromagnetic  Ising model, a vanishingly small external field suffices to 
break this symmetry and direct the relaxation to a unique absorbing  configuration. 

In terms of the order parameters $g$ (the ratio between the number of domains and the lattice area ) and $\left \langle S_{max} \right \rangle/L^2$ 
(the ratio between the size of the largest  cluster and the lattice area)
the scenario is as follows. Except for $p=0$, $g$ is nonzero and so the number of domains grows linearly with the lattice area -- this is the media induced diversity.
However, for all values of $p$ there is a single macroscopic cluster -- the media cluster -- the size of which increases as the media strength $p$  decreases  and
ultimately encompasses the entire lattice  for $p=0$ (explaining this counterintuitive finding is the goal of this paper). 
All other domains are of finite size (typically 3 sites in the thermodynamic limit) but they are so numerous 
that the sum of their sizes grows with $L^ 2$ (see fig. \ref{fig:4}). Therefore, within the perspective that there is a single macroscopic domain we may
say that the system is ordered in the presence of the media regardless of the value of its  strength  $p$.

Although we have restricted our analysis to a single
setting of the control parameters, namely, $F=q=5$,  due mainly to the computational demand of the simulations
of large lattices, our conclusion is valid in general since it provides an intuitive first-principles explanation
for the effect of the media in Axelrod's model, which is not affected by the tuning of the control parameters of the model.

The somewhat trivial  nature of the cultural domains of Axelrod's model in presence of a external media is reminiscent of
the organization of the domains of the extended majority model \cite{Parisi_03,Galam_05b, Peres_10b}, which is
completely determined by the initial random assignment of cultural traits. As in that model, relaxation to an absorbing configuration
is very fast since the random clusters formed at the beginning are practically unaltered  by the update rules of the model.

In summary, we show here that the effect of external global media in Axelrod's model  can be described  much more prosaically  by pointing out that the local homogenizing interactions 
destabilize the domains of agents (usually, isolated sites) which have zero similarity with the media and hence are not affected by it. As a result, decreasing the
media strength and, consequently, increasing the role of the local interactions will favor less fragmented absorbing configurations. Reversely, increasing the media
strength will favor more heterogeneous absorbing configurations. It seems to us that the cause of  the ruckus about the media effect in
Axelrod's model was the failure to realize that the absorbing configurations can already be very fragmented due to the interactions with the  media only.
In fact, the mere inclusion of the data for $p=1$ in figs.\ \ref{fig:1} and \ref{fig:4}  turns the media effect from a surprising phenomenon to a trivial one.

\acknowledgments
This research was supported by The Southern Office of
Aerospace Research and Development (SOARD), Grant No.
FA9550-10-1-0006, and Conselho Nacional de Desenvolvimento
Cient\'{\i}fico e Tecnol\'ogico (CNPq).

\end{document}